\documentclass[preprint,amssymb,numerical,superscriptaddress]{revtex4-1}

\usepackage{subfigure}	% use for side-by-side figures
\usepackage{hyperref}	% use for hypertext links, including those to external documents and URLs
\usepackage{mathtools,tensor}

\newcommand{\eqs}{\begin{eqnarray*}}
\newcommand{\eqf}{\end{eqnarray*}}
\newcommand{\lef}{\left(}
\newcommand{\rig}{\right)}
\newcommand{\mas}{\begin{array}}
\newcommand{\maf}{\end{array}}
\newcommand{\deriv}{\textnormal{d}}

\newcommand{\Deltabd}[1]{\Delta_{\mathsf{bd}_{#1}}}
\newcommand{\e}{\mathrm e}
\newcommand{\EF}{E_{\mathrm F}}
\newcommand{\imu}{\mathrm i}
\newcommand{\kB}{k_{\mathrm B}}

\newcommand{\R}{\mathbf R}
\newcommand{\Rp}{\mathbf R^{\prime}}
\newcommand{\rhogb}{\rho^{\mathsf{GB}}}
\newcommand{\rimp}{\mathbf{r}^{\mathsf{imp}}}
\newcommand{\Sgb}{S^{\mathsf{GB}}}
\newcommand{\Simp}{S^{\mathsf{imp}}}

\newcommand{\vecr}{\mathbf r}
\newcommand{\Vgb}{V^{\mathsf{GB}}}
\newcommand{\Vimp}{V^{\mathsf{imp}}}

\begin{document}

\title{
	Validity criteria for Fermi's golden rule scattering rates applied
	to metallic nanowires
}
\author{Kristof Moors}
\email[E-mail me at: ]{kristof@itf.fys.kuleuven.be}
\affiliation{KU Leuven, Institute for Theoretical Physics, Celestijnenlaan 200D, B-3001 Leuven, Belgium}
\affiliation{Imec, Kapeldreef 75, B-3001 Leuven, Belgium}
\author{Bart Sor\'ee}
\affiliation{Imec, Kapeldreef 75, B-3001 Leuven, Belgium}
\affiliation{University of Antwerp, Physics Department, Groenenborgerlaan 171, B-2020 Antwerpen, Belgium}
\affiliation{KU Leuven, Electrical Engineering (ESAT) Department, Kasteelpark Arenberg 10, B-3001 Leuven, Belgium}
\author{Wim Magnus}
\affiliation{Imec, Kapeldreef 75, B-3001 Leuven, Belgium}
\affiliation{University of Antwerp, Physics Department, Groenenborgerlaan 171, B-2020 Antwerpen, Belgium}

\date{\today}
 
\begin{abstract}
Fermi's golden rule underpins the investigation of mobile carriers
propagating through various solids, being a standard tool to
calculate their scattering rates. As such, it provides a perturbative
estimate under the implicit assumption that the effect of the interaction
Hamiltonian which causes the scattering events is sufficiently small.
To check the validity of this assumption, we present a general
framework to derive simple validity criteria in order to assess whether
the scattering rates can be trusted for the system under consideration,
given its statistical properties such as average size, electron density,
impurity density et cetera. We derive concrete validity criteria for
metallic nanowires with conduction electrons populating a single parabolic
band subjected to different elastic scattering mechanisms: impurities,
grain boundaries and surface roughness.
\end{abstract}

\maketitle

\section{Introduction}
Fermi's golden rule has been applied rather successfully to describe
scattering and obtain the transport properties in various condensed
matter systems.
Ample examples can be found in literature, covering a great variety of
devices and applications such as (conventional) metal-oxide-semiconductor
transistors \cite{abramo1994comparison,mazzoni1999surface,esseni2004modeling,
jin2007modeling,lizzit2014new}, quantum cascade lasers \cite{iotti2001nature}
and nanowire transistors \cite{jin2007modeling2,lenzi2008investigation,
jin2008simulation}, metallic thin films or nanowires \cite{mayadas1970electrical,
moors2014resistivity}, quasi-1D or -2D materials and devices
\cite{pennington2003semiclassical,stauber2007electronic,xu2012impurity,
paussa2013exact,dugaev2013edge,fischetti2013pseudopotential}, as
well as for other applications, e.g. the Hall effect
\cite{sinitsyn2008semiclassical}, spin \cite{piechon2007spin} or thermal
\cite{shi2014nonequilibrium} transport. Being invoked in a straightforward
manner, Fermi's golden rule, however, comes with some limitations and drawbacks,
the most important one being its perturbative nature which essentially
restricts the treatment of all scattering events to the level of second-order
perturbation theory
with respect to the scattering potential $V$.

In order to check whether the perturbative estimate prescribed by
Fermi's golden rule is accurate, the natural thing to do is calculate
the higher-order contributions, compare them to the second-order
scattering rate and verify whether they are indeed negligible.
However, a systematic verification of Fermi's golden rule up to all
orders of the scattering potential is hardly possible for a general
scattering potential representing various scattering agents in a
condensed matter system, such as impurities, phonons, Coulomb
interaction et cetera. Therefore, in most practical cases a
higher-order analysis can only be carried out qualitatively or is
even missing entirely.

To perform a quantitative analysis of the Fermi's golden rule scattering
rates, we present a framework to obtain the higher-order contributions
systematically, leading to validity criteria that can be easily applied.
As these criteria are aimed to be as general as possible, we introduce
an averaging procedure to capture the essential statistical properties
of the scattering potential profiles. Imposing these validity criteria
one can check the validity of transport properties obtained through
Fermi's golden rule scattering rates, without the necessity of comparing
to non-perturbative treatments of scattering potentials
\cite{tevsanovic1986quantum,ness2006quantum,martinez2009perturbative,
oh2012phonon,rideau2012mobility,arenas2015effect}. A non-perturbative
treatment could be too computationally intensive or rely heavily on
other approximations, such that it would be difficult to pinpoint which
of the different approaches is incorrect and for which reason.

In section \ref{sectionFGR} the derivation of Fermi's golden rule and the
higher-order contributions is briefly discussed, as well as the ensemble
averaging procedure for the scattering potentials. Concrete examples
of validity criteria are derived in section~\ref{sectionMetalNW} based
on the third- and fourth-order contributions to Fermi's golden rule for
three types of elastic scattering in metallic nanowires assuming a single
parabolic band for the conduction electrons. In particular, scattering
events due to a single impurity, grain boundaries and surface roughness are
considered. Finally, a discussion of the results and a conclusion are
respectively presented in section \ref{sectionDiscussion} and
\ref{sectionConclusion}.

\section{Fermi's golden rule}
\label{sectionFGR}
We derive Fermi's golden rule by making use of the interaction picture
in which the evolution operator $U(t)$ satisfies the dynamical equation
\begin{align}
   \imu \hbar \frac{ \deriv U\lef t, t_0 \rig}{\deriv t} =
   V (t) U \lef t, t_0 \rig,
\end{align}
with $V$ the scattering potential. This equation can be formally
integrated to obtain $U(t, t_0)$ for all times starting from $t_0$.
Up to second order in $V$, the overlap between an initial state
$\mid \! i \rangle$, having evolved to time $t$, and a final state
$\mid \! f \rangle$ (different from the initial state) is given by
\begin{align}
   C_{fi}(t) & \equiv \langle f \mid U\lef t,t_0 \rig \mid i \rangle.
\end{align}
We now proceed with the standard adiabatic approach, considering
the limit $t_0 \rightarrow -\infty$ and assuming a slow turn-on of
the potential: $V \rightarrow V e^{\eta t}$.
Playing the role of an inverse time scale, $\eta$ is tuned to be in
the regime $t \ll \eta^{-1} \ll t - t_0$, from which the overlap
integral can be obtained to any order in $V$,
\begin{align} \label{eq:Cfit}
   &C_{fi}(t) \\ \notag
   &\; = \frac{e^{[ \eta - \imu (E_i - E_f)/\hbar ] t}}
            {E_i - E_f + \imu \hbar\eta}
       \lef
           \langle f \mid V \mid i \rangle + \sum_\alpha \mkern-3mu
           \int \mkern-5mu \deriv E \;
           \frac{\deriv n_\alpha}{\deriv E}
           \frac{\langle f \mid V \mid E, \alpha \rangle \langle E,
                 \alpha \mid V \mid i \rangle}
                {E_i - E + \imu \hbar \eta}
           + \ldots
       \rig.
\end{align}
The intermediate states can be labeled by an energy eigenvalue $E$
and a (sub)band index $\alpha$ to lift the remaining degeneracy.
The density of states for each (sub)band is denoted by
$\deriv n_\alpha / \deriv E$. Eq.~\ref{eq:Cfit} can be used to
obtain the scattering rate $1/\tau_{i\rightarrow f}$ which
describes the transition $\mid \! i \rangle \to \; \mid \! f \rangle$,
by taking the time derivative of the absolute value squared of the
overlap between the two states. In the special case of a
one-dimensional conductor of length $L_z$, for which the entire phase
space is captured by a one-dimensional wave vector $k$ along
the transport direction $z$, we get
\begin{align} \label{eq:rate}
    \frac{1}{\tau_{i\rightarrow f}} &\equiv \frac{\deriv}{\deriv t}
	\left| C_{fi} (t) \right|^2 \\ \notag
    &\approx \frac{2\pi}{\hbar} \delta \lef E_i - E_f \rig
	\left| \langle f \mid V \mid i \rangle + \sum_\alpha
	\frac{L_z}{2\pi} \mkern-2mu \int \mkern-4mu \deriv k \;
	\frac{\langle f \mid V \mid k, \alpha \rangle
	\langle k, \alpha \mid V \mid i \rangle}
	{E_i - E_\alpha(k) + \imu\hbar\eta} \right|^2,
\end{align}
where the energy-wave vector relation $E_\alpha(k)$ has been introduced.
The resulting scattering rate can be interpreted as the total rate
incorporating both a direct transition from initial to final state
and indirect transitions with one or more intermediate states.
Next, we evaluate the integral in Eq.~\ref{eq:rate} as a contour
integral, the integrand having poles $k_\alpha$ that satisfy
\begin{align} \label{eqPoles}
E_i - E_\alpha(k_\alpha) + \imu\hbar\eta = 0.
\end{align}
As the behavior of the $k$-dependent matrix elements determines how
the complex contour is to be closed, we proceed by writing out
explicitly the matrix elements,
\begin{align} \label{matrixEl2ndOrder}
   & \langle f \mid V \mid k, \alpha \rangle \langle k, \alpha
     \mid V \mid i \rangle \\ \notag
   &\; = \mkern-3mu \int \mkern-5mu \deriv^2 R \;
         \psi_f^*(\R) \psi_\alpha(\R)
         \mkern-3mu \int \mkern-5mu \deriv^2 R' \;
         \psi_\alpha^*(\Rp) \psi_i(\Rp)
         \mkern-11mu \int \limits_{-L_z/2}^{+L_z/2}
         \mkern-13mu \deriv z \; V(\R, z)
         \mkern-11mu \int \limits_{-L_z/2}^{+L_z/2}
         \mkern-13mu \deriv z' \; V(\Rp,z') \;
         \frac{e^{- \imu (k_f - k) z - \imu (k - k_i)z'}}{L_z^2},
\end{align}
where the wave functions of the initial and final states as well as of the
intermediate states factorize into plane waves along the transport direction
$z$ and envelope functions $\psi(\R)$ for the transverse directions
$\R \equiv (x,y)$.
The half plane for which the matrix element product rapidly tends to zero
when $k$ is on a semi-circle with increasing radius, is determined by the
sign of $z'-z$, as long as the $z, z'$-dependence of $V$ is such that it
does not prevent the exponential decrease due to the plane wave solution.
The integration over $k$ gives
\begin{align} \label{indirectTrans}
    &\mkern-3mu \frac{L_z}{2 \pi} \int\limits_{-\infty}^{+\infty} \mkern-7mu
	\deriv k \; \frac{e^{- \imu (k_f - k) z - \imu (k - k_i) z'}}
	{E_i - E_\alpha(k) + \imu\hbar\eta} \\ \notag
    &\; = \imu L_z \; \sum_{k_\alpha} \lim\limits_{k \rightarrow k_\alpha}
	\frac{e^{ - \imu (k_f - k) z - \imu (k - k_i) z'}}{E_i - E_\alpha(k)
	+ \imu\hbar\eta} \times
	\left\{ \begin{matrix}
	    \lef k - k_\alpha \rig \qquad
	    \textnormal{if} \quad \mathcal{I}\lef k_\alpha \rig > 0 \quad
	    \textnormal{and} \quad z'-z < 0 \\
	    \lef k_\alpha - k \rig \qquad \textnormal{if} \quad
	    \mathcal{I}\lef k_\alpha \rig < 0 \quad \textnormal{and}
	    \quad z'-z > 0
	\end{matrix} \right. .
\end{align}
If the conditions inside the parentheses are not met, the contribution
from the pole is zero. This procedure can be repeated for all higher-order
contributions, leading to simple diagrammatic rules, as summarized in
appendix \ref{appendixFeynman}.
The derivation can also be done more formally to any order in $V$ by
using the T-matrix \cite{zubarev1996statistical}.
\begin{figure}[tb]
   \begin{center}
      \subfigure[]{
      \includegraphics[width=0.4\linewidth]{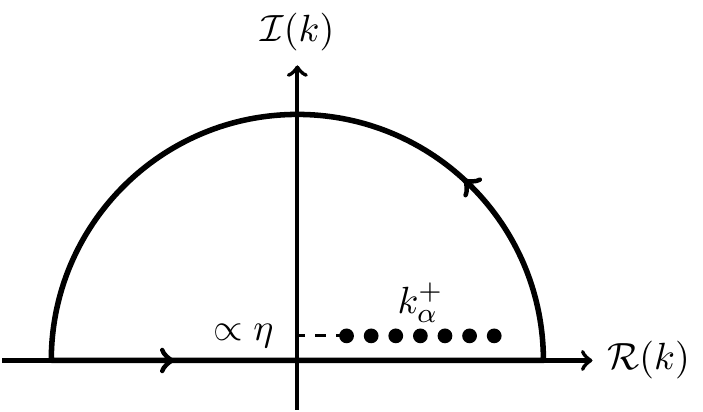}}
      \subfigure[]{
      \includegraphics[width=0.4\linewidth]{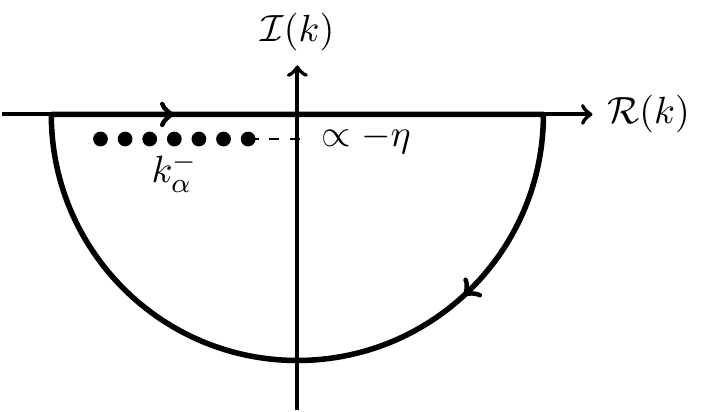}}
   \end{center}
\caption{The poles $k_\alpha$ that are picked up by the contour integration
	 as performed in Eq.~\ref{indirectTrans} are shown for an energy-momentum
	 relation symmetric around $k=0$ (e.g. the effective mass model discussed
	 in section~\ref{sectionMetalNW}. The imaginary part is negative/positive
	 for negative/positive $k_\alpha$ and proportional to the slow turn-on $\eta$.}
\label{finiteDomain}
\end{figure}

\subsection{Ensemble averaging}
\label{sectionEnsAv}
In many cases of interest, straightforward application of Fermi's golden
rule suffers from an incomplete knowledge of the scattering potential.
In practice, this is reflected in the lack of detailed information
regarding the spatial or orientational distribution of the scattering
sources in realistic, non-ideal condensed matter systems. The brute
force solution to this problem would amount to repeating all relevant
simulations for a huge number of potential profiles, while applying
statistics to the outcome. However, as such a procedure would lead to
unreasonably high computation times when it comes to derive general,
but simple validity criteria for Fermi's golden rule, it becomes
paramount to construct a tractable, analytical expression for a generic
scattering rate that is averaged over an ensemble of potentials,
all representing a particular configuration of the relevant scattering
sources. Similar averaging techniques have been applied before to
various scattering models, e.g. the random phase approximation for
impurities, Gaussian or exponential statistics for Ando's surface
roughness model \cite{ando1982electronic} and a Gaussian distribution
of grain boundaries in the Mayadas-Shatzkes model
\cite{mayadas1970electrical}, but here we present the averaging
procedure in a formal way in view to the envisaged validity criteria.
In this light, we need to deal with ensemble averages of matrix
elements and products thereof involving initial an final states
$\mid \! i \rangle, \, \mid \! f \rangle$ as well as intermediate
states $\mid \! k, \alpha \rangle$. We introduce the following notation
to represent matrix elements explicitly as functionals of the potential
profile $V(\mathbf{r})$,
\begin{equation}
   \langle i \mid V \mid f \rangle \equiv \mathcal{M}_{if} [V], \qquad
   \langle i \mid V \mid k,\alpha \rangle
   \langle k,\alpha \mid V \mid f \rangle
   \equiv \mathcal{M}_{i \alpha f} [V], \qquad \ldots,
\end{equation}
Taking the ensemble average of a matrix element is then accomplished by
the following functional integration,
\begin{equation} \label{eqAveraging}
   \left< \langle i \mid V \mid f \rangle \right>_{V} \equiv
   \int \mkern-2mu \delta V(\mathbf{r}) \; g[V (\mathbf{r})] \;
   \mathcal{M}_{if} \left[ V(\mathbf{r}) \right],
\end{equation}
where $g[V(\mathbf{r})]$, the distribution functional describing
the ensemble of relevant potentials, is properly normalized
according to
\begin{equation}
   \int \mkern-2mu \delta V(\mathbf{r}) \; g[V (\mathbf{r})] = 1.
\end{equation}
Generally depending on the type of potentials we need to consider (see
sections \ref{sectionImp}, \ref{sectionGB} and \ref{sectionSR}),
$g[V(\mathbf{r})]$
will often be replaced by an ordinary distribution function.

As an example, we quote the averaged scattering rate $1/\tau_{i\rightarrow f}$
up to lowest order in $V$ with the notation introduced in this section,
\begin{align}
   \left\langle \frac{1}{\tau_{i \rightarrow f}} \right\rangle_V &=
   \frac{2\pi}{\hbar} \delta \lef E_i - E_f \rig \left\langle \left|
   \mathcal{M}_{if} \right|^2 \right\rangle_V, \\
   \left\langle \left|
   \mathcal{M}_{if} \right|^2 \right\rangle_V &= \int \mkern-2mu
   \delta V(\mathbf{r}) \; g[V (\mathbf{r})] \;
   \left| \mathcal{M}_{if} \left[ V(\mathbf{r}) \right] \right|^2.
\end{align}
\section{Metallic nanowire}
\label{sectionMetalNW}
We will develop validity criteria for application of Fermi's golden rule
to scattering rates of electrons in a single parabolic band in a metallic
nanowire. The single-electron eigenstates are denoted by
$\mid \! k, \mathbf{n} \rangle$ where $\mathbf{n}$ is a shorthand notation
indicating the subband indices $(n_x, n_y)$ that label the electron subbands
along the two transverse directions ($x$ and $y$) and $k$ is the wave vector
along the transport direction $z$.
Dealing with metallic wires and assuming that the Fermi energy (relative
to the bottom of the conduction band) is large compared to $\kB T$
(typically a safe assumption), we may assert that only states with
energies equal to the Fermi energy $\EF$ participate in elastic
scattering processes.
Adopting further the effective mass approximation, we infer parabolic
dispersion relations for the conduction band and its subbands, as well as
linear group velocities,
\begin{equation}
   E_\alpha (k) = E_\alpha^0 + B_\alpha k^2, \qquad
   \left.
         \frac{\deriv E_\alpha}{\deriv k}
   \right|_{E=\EF}
   \mkern-10mu = \theta \left[ B_\alpha \lef \EF - E_\alpha^0 \rig \right]
   2 \sqrt{B_\alpha (\EF - E_\alpha^0)}.
\end{equation}
Inserting the dispersion relation into Eq.~\ref{indirectTrans}, we obtain
\begin{align}
   \frac{L_z}{2 \pi}
   & \int\limits_{-\infty}^{+\infty} \mkern-7mu \deriv k \;
     \frac{\langle f \mid V \mid k, \alpha \rangle 
           \langle k, \alpha \mid V \mid i \rangle}
          {E_i - E_\alpha(k) + \imu \hbar\eta}
     = \frac{-\imu L_z}{2 B_\alpha k_\alpha^+}
       \overline{\mathcal{M}_{f \alpha}
                 \mathcal{M}_{\alpha i}} \\ \notag
   \; \equiv \,
   & \frac{-\imu}{2 B_\alpha k_\alpha^+ L_z}
     \mkern-3mu \int \mkern-5mu \deriv \R \;
     \psi_f^*(\R) \psi_\alpha(\R)
     \mkern -3mu \int \mkern-5mu \deriv \Rp \;
     \psi_\alpha^*(\Rp) \psi_i(\Rp) \\ \notag
    & \times \mkern-12mu \int\limits_{-L_z/2}^{+L_z/2}
    \mkern-14mu \deriv z \;
    V(\R,z) \mkern-12mu \int\limits_{-L_z/2}^{+L_z/2}
    \mkern-14mu \deriv z' \; V(\Rp,z') \;
    \e^{-\imu k_f z - \imu k_\alpha^+ |z'-z| - \imu k_i z'},
\end{align}
where $\R$ and $\Rp$ are transverse position vectors and
$k_\alpha^+ \equiv \sqrt{\lef \EF - E^0_\alpha \rig/B_\alpha}$ denotes
the positive pole of subband $\alpha$. The line above the product of
matrix elements $\overline{\mathcal{M}_{f \alpha} \mathcal{M}_{\alpha i}}$
denotes the replacement of the difference of position coordinates of
the intermediate state wave functions $z-z'$ by its absolute value and
the insertion of the pole $k_\alpha^+$ into the equation. This procedure
is performed in order to keep the full integration domain for $z$ and $z'$
from $-L_z/2$ to $+L_z/2$ while only inserting a single pole into the
expression, preventing the splitting of the integration domain according
to Eq.~\ref{indirectTrans}. This replacement and elimination of one of
two poles can be performed for each pair of coordinates along the
transport direction that belongs to an intermediate state and is
presented in the diagrammatic rules in appendix \ref{appendixFeynman}.

Carrying out a contour integration in accordance with Eq.~\ref{indirectTrans}
and performing the ensemble averaging as explained in section \ref{sectionEnsAv},
we get the following validity criteria involving the third- and
fourth-order contributions arising from the scattering potential $V$,
\begin{widetext}
\begin{align}
   \label{crit1}
   \frac{\left|
               \left\langle
                    2 \mathcal{R}
                    \left[
                          \sum\limits_{\alpha}
                          \lef \frac{- \imu L_z}{2 B_\alpha k_\alpha^+} \rig
                          \overline{\mathcal{M}_{f\alpha i}} \mathcal{M}_{if}
                    \right]
               \right\rangle_V
         \right|}
         {\left\langle
               \left| \mathcal{M}_{if} \right|^2
          \right\rangle_V}
          & \ll 1, \\
   \label{crit2}
   \frac{\left|
               \left\langle
                    \left|
                          \sum\limits_{\alpha}
                          \lef \frac{- \imu L_z}{2 B_\alpha k_\alpha^+} \rig
                          \overline{\mathcal{M}_{f\alpha i}} \right|^2 +
                          2 \mathcal{R}
                          \left[
                                \sum\limits_{\alpha,\alpha'}
                                \lef
                                    \frac{- \imu L_z}{2 B_\alpha k_\alpha^+}
                                \rig
                                \lef
                                    \frac{- \imu L_z}
                                         {2 B_{\alpha'} k_{\alpha'}^+}
                                \rig
                                \overline{\mathcal{M}_{f\alpha\alpha' i}}
                                \mathcal{M}_{if}
                          \right]
               \right\rangle_V
         \right|}
         {\left\langle \left| \mathcal{M}_{if} \right|^2 \right\rangle_V}
         & \ll 1,
\end{align}
\end{widetext}
with all the matrix elements being evaluated at $k,k'=k_\alpha^+,k_{\alpha'}^+$.
From hereof we will refer to the $n$-th order contribution as $\mathcal{O}(V^n)$,
thus implying that Eqs.~\ref{crit1}-\ref{crit2} reduce to
$|\mathcal{O}(V^3)/\mathcal{O}(V^2)| \ll 1$,
$|\mathcal{O}(V^4)/\mathcal{O}(V^2)| \ll 1$ in short.
Although the lowest higher-order criteria are expected to provide a
reliable validity assessment for all higher-order corrections,
some higher-order contributions might cancel out exactly or be
relatively small due to the properties of the scattering potential.
This appears to be the case for the third-order contribution related to
scattering events treated in sections~\ref{sectionImp}-\ref{sectionGB},
\ref{sectionSR}, the fourth-order validity criterion typically giving
rise to a much stronger constraint on the scattering potential size.

Below we consider three examples of elastic scattering represented by an
appropriate scattering potential for a localized impurity, grain boundaries
and surface roughness.

\subsection{Single impurity}
\label{sectionImp}
\begin{figure}[tb]
   \begin{center}
      \includegraphics[width=0.5\linewidth]{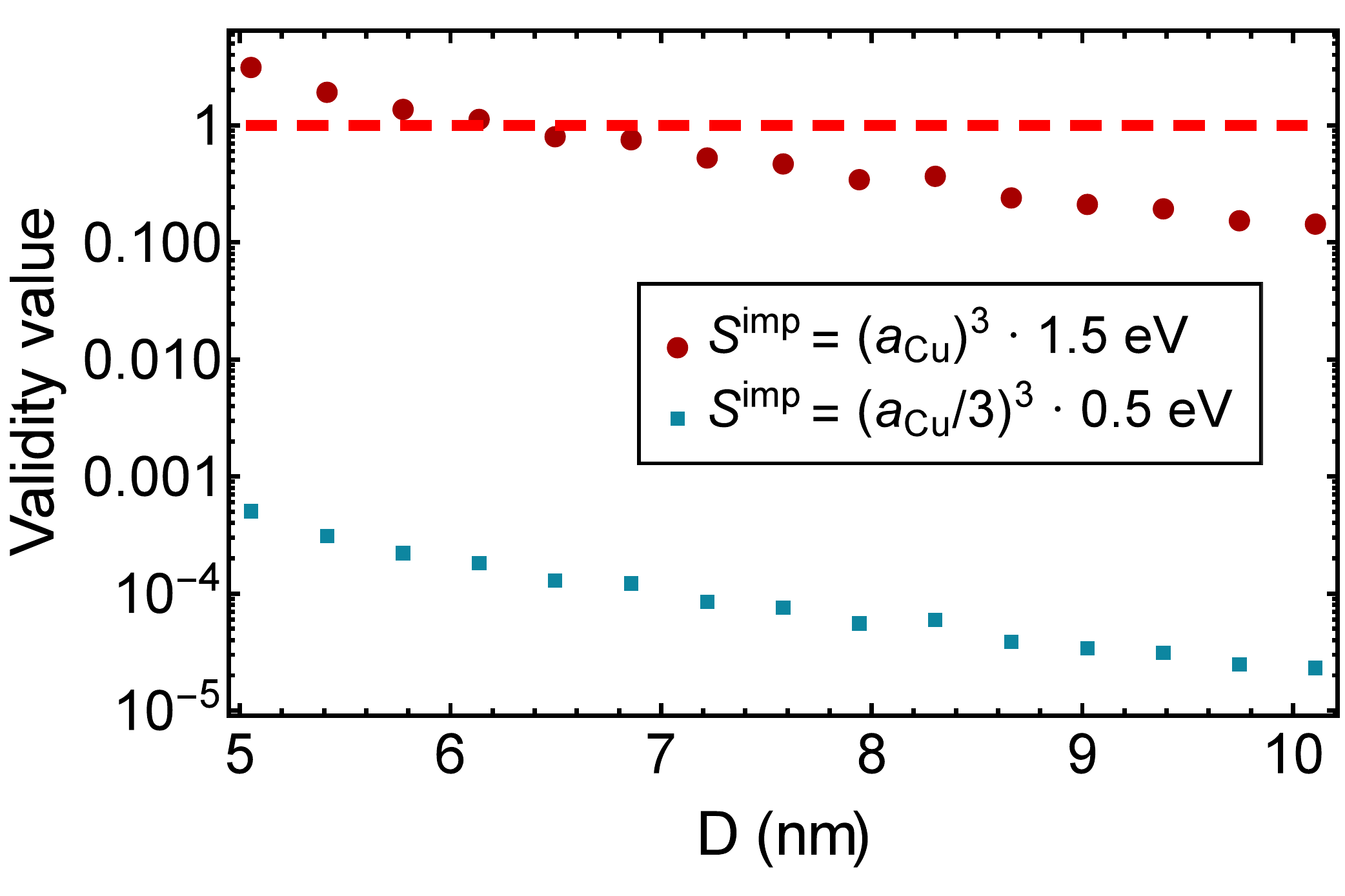}
   \end{center}
   \caption{The validity criteria for scattering by a single impurity
	    are evaluated for nanowires with square cross section
	    ($D\equiv L_x = L_y$) for two impurity strengths:
	    $\Simp = \lef a_\textnormal{Cu} \rig^3 \cdot 1.5$~eV
	    and $\Simp = \lef a_\textnormal{Cu}/3 \rig^3 \cdot 0.5$~eV.
	    The evaluation is performed for copper nanowires,
	    assuming a lattice constant $a_\textnormal{Cu} \approx 0.361$~nm
	    and an electron density: $n_{\mathsf e} \approx
	    8.47 \times 10^{22}$~cm${}^{-3}$.}
   \label{FigImpValidity}
\end{figure}

For the sake of simplicity, we consider a single impurity at a random
position $\rimp$, the impurity potential taking the form of a delta
function with strength $\Simp$,
\begin{equation}
   \Vimp(\vecr) \equiv \Simp \, \delta(\vecr - \rimp).
\end{equation}
The unperturbed Hamiltonian describing a quasi-free electron (with
effective mass $m^*$) in an ideal, boxed wire with zero potential
inside the wire and infinite potential outside, and its transverse
wave functions are given by:
\begin{align}
   H_0 (\vecr)
   & \equiv -\frac{\hbar^2 \nabla^2}{2 m^*} +
     \begin{cases}
        0       & \textnormal{if} \quad 0 \leqslant x, y \leqslant L_x, L_y \\
	+\infty & \textnormal{else}
     \end{cases}, \\
   \psi_\alpha(x, y)
   & = \lef 2 / \sqrt{L_x L_y} \rig \sin \lef n_{\alpha \, x} \pi x / L_x \rig
       \sin \lef n_{\alpha \, y} \pi y / L_y \rig,
       \quad n_{\alpha \, x}, n_{\alpha \, y} = 1, 2, 3, \ldots
\end{align}
Next, assuming a uniform impurity distribution, we may
replace the averaging functional integral by
\begin{equation}
   \int \mkern-2mu \delta V(\mathbf{r}) \; g[V (\mathbf{r})]
   = \frac{1}{L_x L_y L_z} \int\limits_0^{L_x}
      \mkern-3mu \deriv x^\textnormal{imp}
      \mkern-3mu \int\limits_0^{L_y}
      \mkern-3mu \deriv y^\textnormal{imp}
      \mkern-11mu \int\limits_{-L_z/2}^{+L_z/2}
      \mkern-13mu \deriv z^\textnormal{imp},
\end{equation}
in order to compute the required averages occurring in the criteria
formulated in Eq.~\ref{crit1}-\ref{crit2}:
\begin{align}
   & \mathcal{O} \lef V^2 \rig =
     \lef \frac{\Simp}{L_x L_y L_z}\rig^2 C_{i f}^\textnormal{imp},
     \qquad \mathcal{O} \lef V^3 \rig = 0, \\
   & \mathcal{O} \lef V^4 \rig = 
     \sum_{\alpha, \alpha'} \lef \frac{L_z}{2 B_\alpha k_\alpha^+} \rig
     \lef \frac{L_z}{2 B_{\alpha'} k_{\alpha'}^+} \rig
     \lef \frac{\Simp}{L_x L_y L_z} \rig^4
     C_{i f \alpha \alpha'}^\textnormal{imp}.
\end{align}
$C_{i f}^\textnormal{imp.},C_{i f \alpha \alpha'}^\textnormal{imp.}$
are positive constants of the order of one arising from the
wave function parts associated with the transverse directions.
Ignoring the factors that are of order one, we get the following
validity criterion for the fourth-order contribution,
\begin{align} \label{ImpValidityEq}
    \left| \sum_{\alpha} \lef \frac{L_z}{2 B_\alpha k_\alpha^+} \rig
    \lef\frac{S^\textnormal{imp.}}{L_x L_y L_z}\rig \right|^2 \ll 1,
\end{align}
while the third order gives no constraint.
The validity check boils down to the comparison of two quantities
having the dimension of a volume times an energy, the first one
being the impurity strength and the second one arising from the
subband density of states at the Fermi level.
Note that the criterion is independent of the wire length when
the impurity size does not scale with the wire length.
The form of the initial and final state wave functions as well as
the intermediate states only influences the details of ignored
factors that are of order one, the effect being minimal because
the impurity position is averaged over the whole wire volume, which
smears out any possibly larger effect. The validity criterion
mentioned in Eq.~\ref{ImpValidityEq} is evaluated for different
nanowire cross sections and impurity strengths, as shown in
Fig.~\ref{FigImpValidity}.

\subsection{Grain boundaries}
\label{sectionGB}
Similarly, we may characterize scattering due to grain boundaries by a
potential that consists of a sum of Dirac delta functions centered around
various axial positions along the wire, representing barrier planes
oriented perpendicularly to the transport direction,
\begin{equation}
   \Vgb(\vecr) \equiv \sum_{j=1}^N \Sgb \delta \lef z - z_j \rig,
\end{equation}
which, essentially, is borrowed from the grain boundary potential proposed
by Mayadas and Shatzkes \cite{mayadas1970electrical}. Adopting the
Mayadas-Shatzkes model, we further assume that all grain boundary
planes are uniformly distributed, while neglecting any correlations.
The latter are expected to be relevant only if the number of boundary
planes is relatively small or if the plane positions were to form a periodic
array (thereby enabling the occurrence of resonant tunneling), neither of
which is the case for realistic, metallic nanowires. As a consequence, the
functional integration may be reduced to an ordinary, multiple integral:
\begin{equation}
   \int \mkern-2mu \delta V(\vecr) \; g[V (\vecr)] \,
   (\ldots) =
   \frac{1}{L_z^N} \prod_{j=1}^N \int\limits_{-L_z/2}^{+L_z/2}
   \mkern-15mu \deriv z_j \, (\ldots) .
\end{equation}
We obtain the following contributions for the scattering rates:
\begin{align}
	\mathcal{O} \lef V^2 \rig &\approx N \lef \frac{\Sgb}{L_z} \rig^2
		\delta_{\mathbf{n}_i, \mathbf{n}_f} \delta_{k_i, -k_f}, \\
	\mathcal{O} \lef V^3 \rig &\approx - \frac{N!}{(N-2)!}
		\frac{1}{B_i \left|k_i \right|^2} \lef \frac{\Sgb}{L_z} \rig^3
		\delta_{\mathbf{n}_i, \mathbf{n}_f} \delta_{k_i, -k_f}, \\
	\mathcal{O} \lef V^4 \rig &\approx \left[ \frac{N!}{(N-4)!}
		\frac{1}{\lef 2 B_i \left| k_i \right|^2 \rig^2} - \frac{N!}{(N-2)!}
		\frac{1}{\lef 2 B_i \left| k_i \right|/L_z \rig^2} \right]
		\lef \frac{\Sgb}{L_z} \rig^4 \delta_{\mathbf{n}_i, \mathbf{n}_f}
		\delta_{k_i, -k_f},
\end{align}
with $\delta_{a,b}$ being a Kronecker delta. Considering the limits
$1/kL_z \ll 1$ and $N \gg 1$, while keeping all contributions of
different orders in $k L_z/N$, we may formulate the validity criteria
related to grain boundary scattering as follows:
\begin{equation} \label{critGB}
   \left| \frac{\rhogb \Sgb}{B_i \left | k_i \right|^2} \right| \ll 1,
   \qquad
   \left|
         N \lef \frac{\rhogb \Sgb}{2 B_i \left| k_i \right|^2} \rig^2 - 
         N^2 \lef \frac{\Sgb}{2 B_i \left| k_i \right|} \rig^2
   \right| \ll 1,
\end{equation}
where $\rhogb \equiv N/L_z$ denotes the grain boundary density.
As a result, the second inequality provides the stronger constraint,
as can be observed in Fig.~\ref{FigGBValidity}). Typically, the
strongest criteria are found to involve the even powers of $V$,
the odd powers appearing in the cross terms and, hence, being
reduced stronger under averaging. The criteria for grain boundary
scattering also depend on the initial state. The dependence on $k_i$
shows that it suffices to check the validity criteria for the lowest
appearing $k_i$ in a metallic nanowire, giving rise to a maximal
ratio in Eq.~\ref{critGB} and providing an upper bound for all $k_i$.
Note that the second inequality in Eq.~\ref{critGB} is length dependent
through its dependence on $N$, even though the lowest-order
contribution leads to length independent transport properties
when a constant grain boundary density is assumed.

For the single impurity in the previous section, one should bear
in mind that the transport properties depend on the wire length,
since the effect of a single impurity diminishes with increasing
wire length. It seems impossible to make both the transport
properties and the validity criteria length independent.
For the sake of comparison, a non-perturbative treatment of grain
boundary scattering is presented in the following subsection.

\subsubsection{Comparison with non-perturbative solution}
\begin{figure}[tb]
\begin{center}
\includegraphics[width=0.9\linewidth]{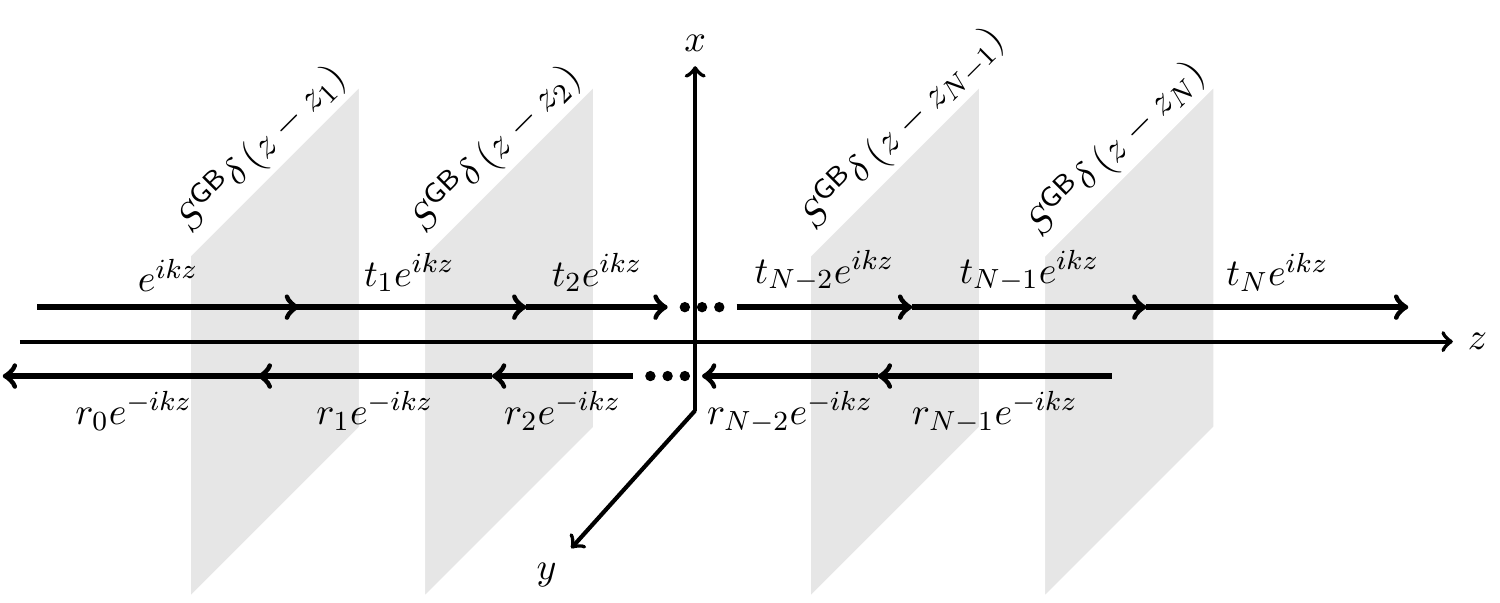} 
\end{center}
\caption{A plane wave with wave vector $k {\mathbf e}_z$ is injected from
	the left and propagates through $N$ grain boundary planes at the positions
	$z_1$, $z_2$, \ldots, $z_N$ with barrier strength $\Sgb$. The total
	transmission probability is given by $|t_N|^2 = 1 - |r_0|^2$.}
   \label{figReflTrans}
\end{figure}
\begin{figure}[tb]i
   \begin{center}
      \subfigure[\ NW1]{
      \includegraphics[width=0.4\linewidth]{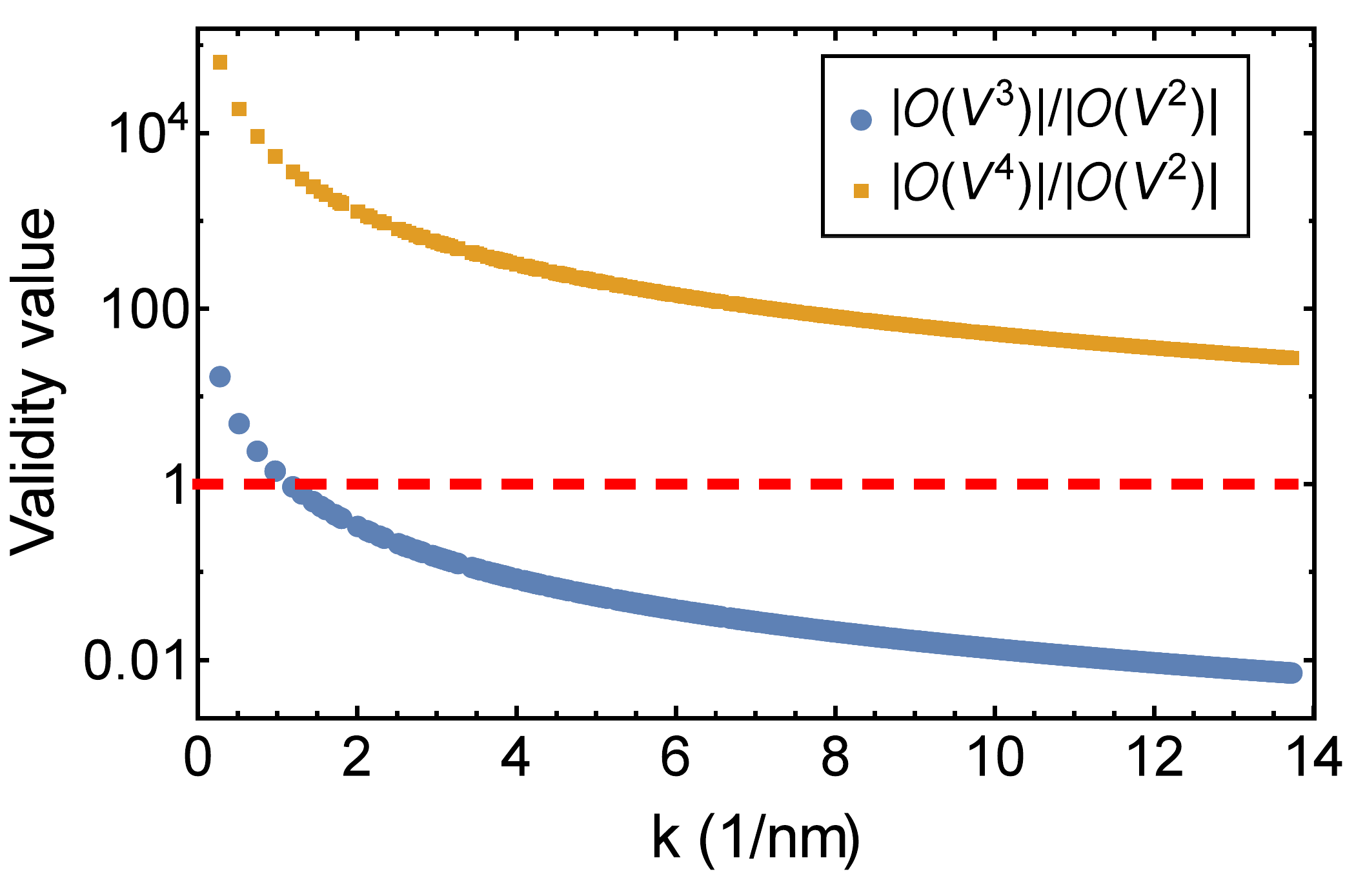}}
      \subfigure[\ NW1 (Non-perturbative)]{
      \includegraphics[width=0.4\linewidth]{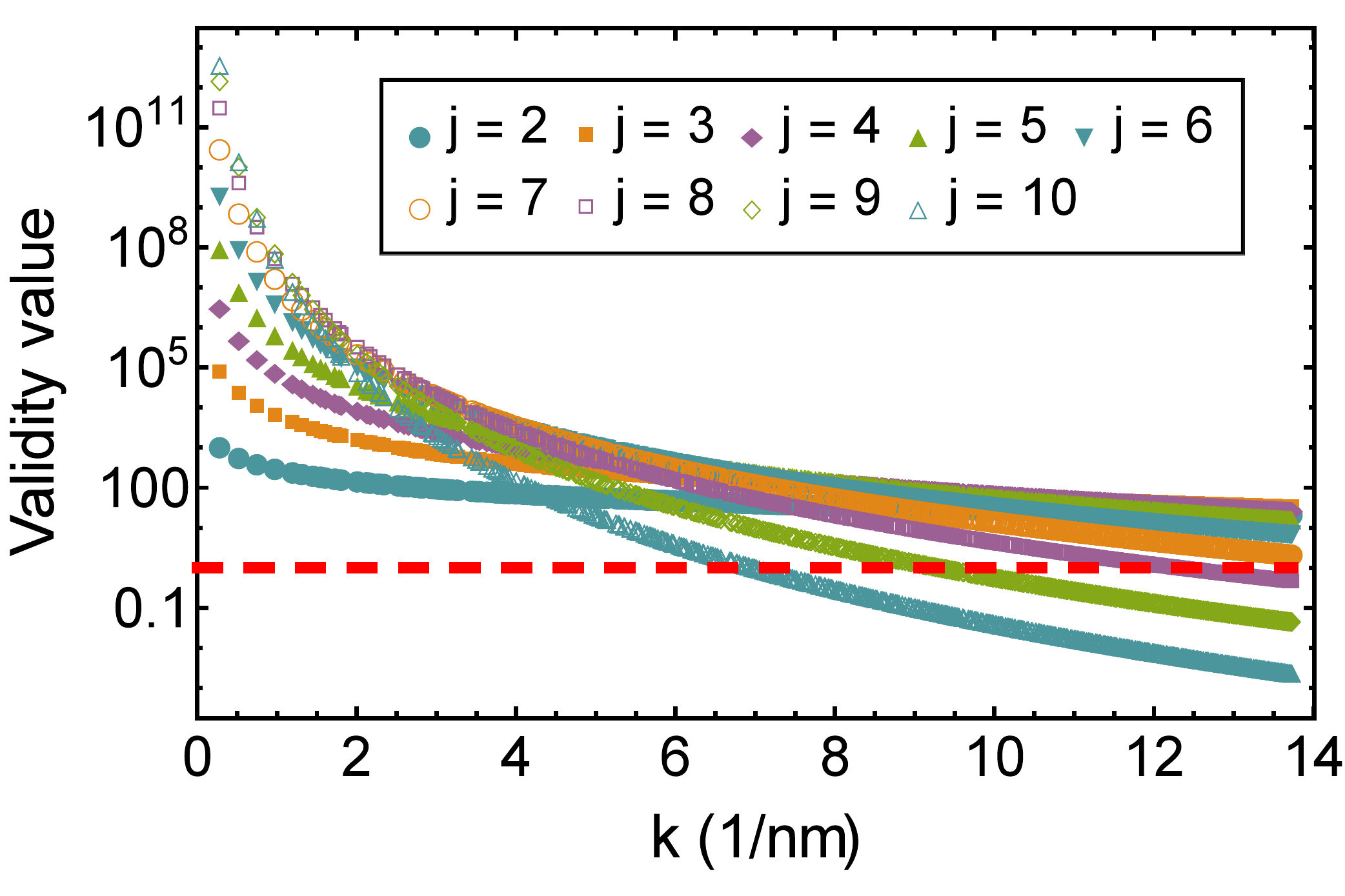}}
      \subfigure[\ NW2]{
      \includegraphics[width=0.4\linewidth]{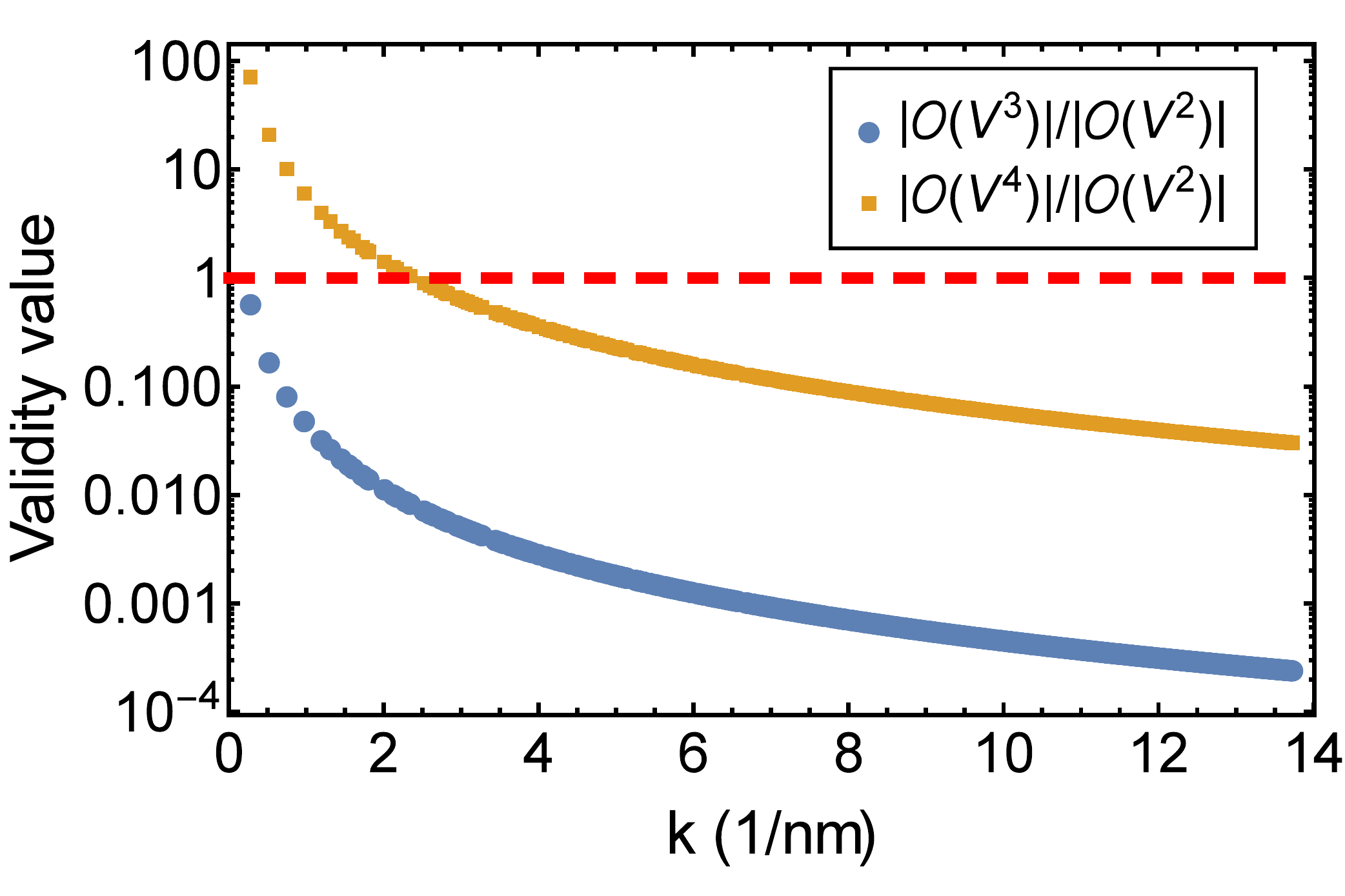}}
      \subfigure[\ NW2 (Non-perturbative)]{
      \includegraphics[width=0.4\linewidth]{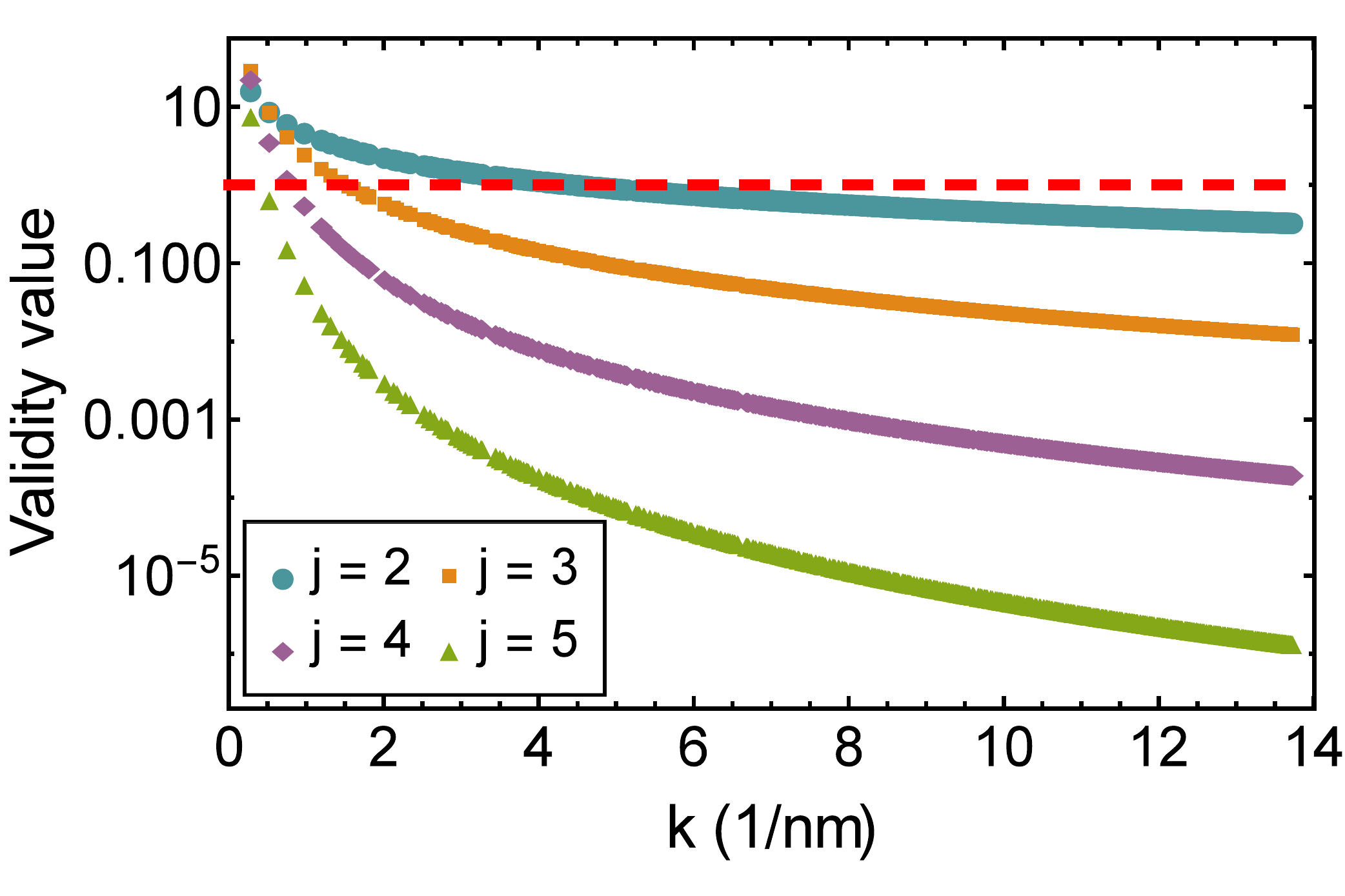}}
   \end{center}
   \caption{Validity criteria are evaluated for two nanowires with length
	    $L_z = 100$~nm, width and height $L_x = L_y \approx 10$~nm,
	    lattice constant $a_\textnormal{Cu} \approx 0.361$~nm and
	    electron density of copper:
	    $n_{\mathsf e} \approx 8.469\cdot 10^{22}$~cm${}^{-3}$.
	    The grain boundary parameters are (NW1)
	    $\Sgb = a_\textnormal{Cu}\cdot 1.5$~eV, $N = 10$ (NW2)
	    $\Sgb = a_\textnormal{Cu}/5 \cdot 0.5$~eV, $N = 5$.}
   \label{FigGBValidity}
\end{figure}

Below, we compare the scattering rates obtained with Fermi's golden rule
with the criteria obtained non-perturbatively by calculating the exact
reflection and transmission coefficients, as depicted in Fig.~\ref{figReflTrans}.
We might expect the non-perturbative solution to diverge from the
golden rule solution when the perturbative analysis is to break down
according to the validity criterion. Upon invoking the random phase
approximation for the positions of the grain boundaries, the transmission
coefficient is given by:
\begin{equation}
   T = | t_N |^2 = 
   \left|
         1 + \sum_{j=1}^N \frac{N!}{(N-j)!j!}
         \lef -\imu \frac{\Sgb}{2 B k} \rig^j
   \right|^{-2}.
\end{equation}
Up to lowest order, the grain boundary scattering rate obtained by
Fermi's golden rule is proportional to $\lef \Sgb \rig^2$. If it is
to agree with the transmission coefficient given above, one needs
to warrant that the contributions to the coefficient that correspond
to higher-order terms in $\Sgb$, be negligible compared to the
second-order contribution, leading to the following constraints:
\begin{equation}
   \left|
         \frac{(N-1)!}{(N-j)!j!} \lef \frac{\Sgb}{2 B k} \rig^{j-1}
   \right|
   \ll 1 \quad \textnormal{for} \; j = 2, \ldots, N.
\end{equation}
The above constraints agree well with the criteria in Eq.~\ref{critGB}
(see Fig.~\ref{FigGBValidity}) and are also affected by the wire
length through the $N$ dependency.
Hence, the validity criteria obtained from the higher-order
(especially fourth-order) corrections to Fermi's golden rule are
confirmed to provide useful validity constraints.
Moreover, they are definitely of interest for cases where
non-perturbative treatments, as derived here for grain boundary
scattering, are difficult or impossible to perform.

\subsection{Surface roughness}
\label{sectionSR}
\begin{figure}[htb]
   \begin{center}
      \subfigure[]{\includegraphics[width=0.4\linewidth]{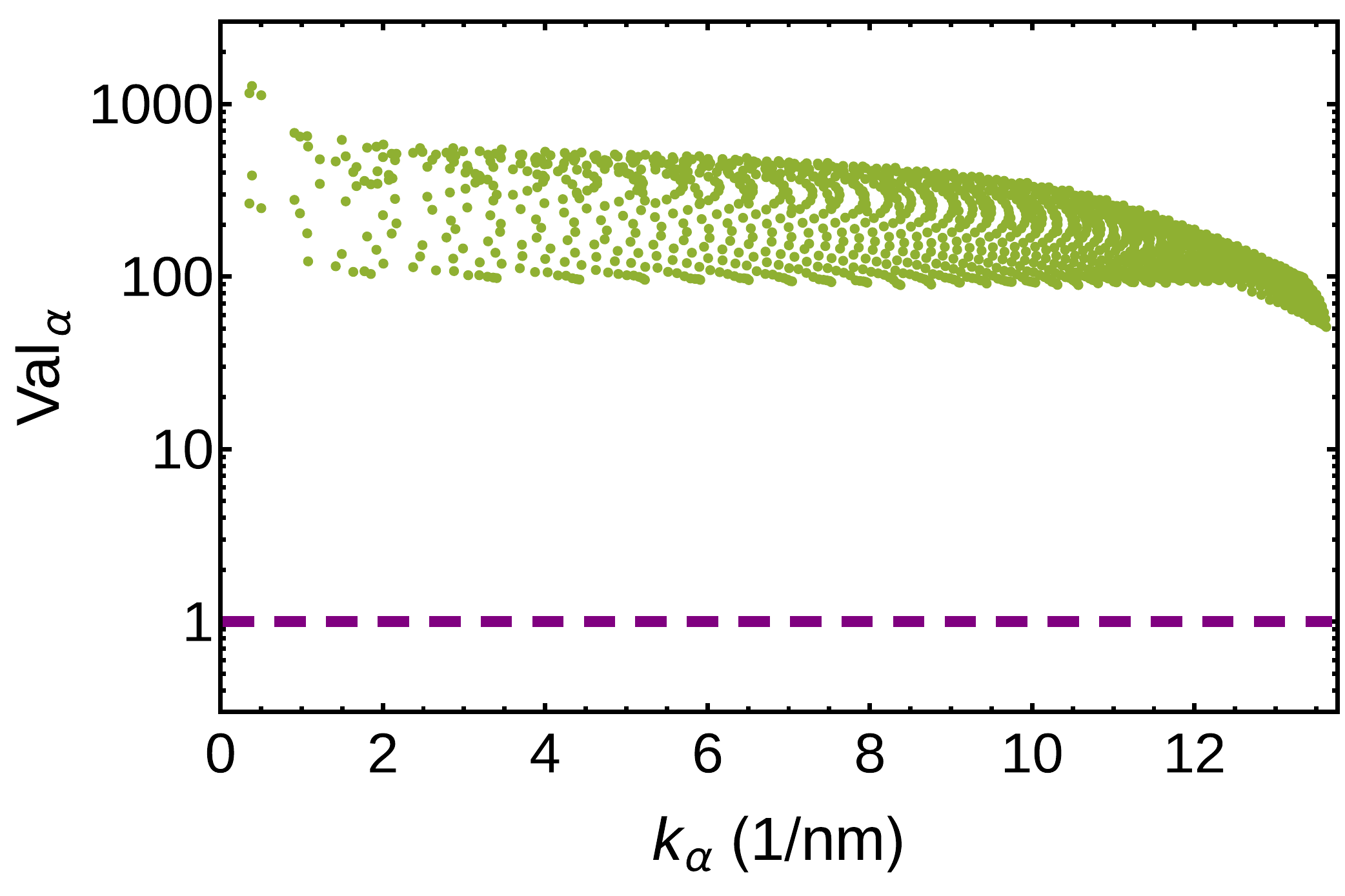}}
      \subfigure[]{\includegraphics[width=0.4\linewidth]{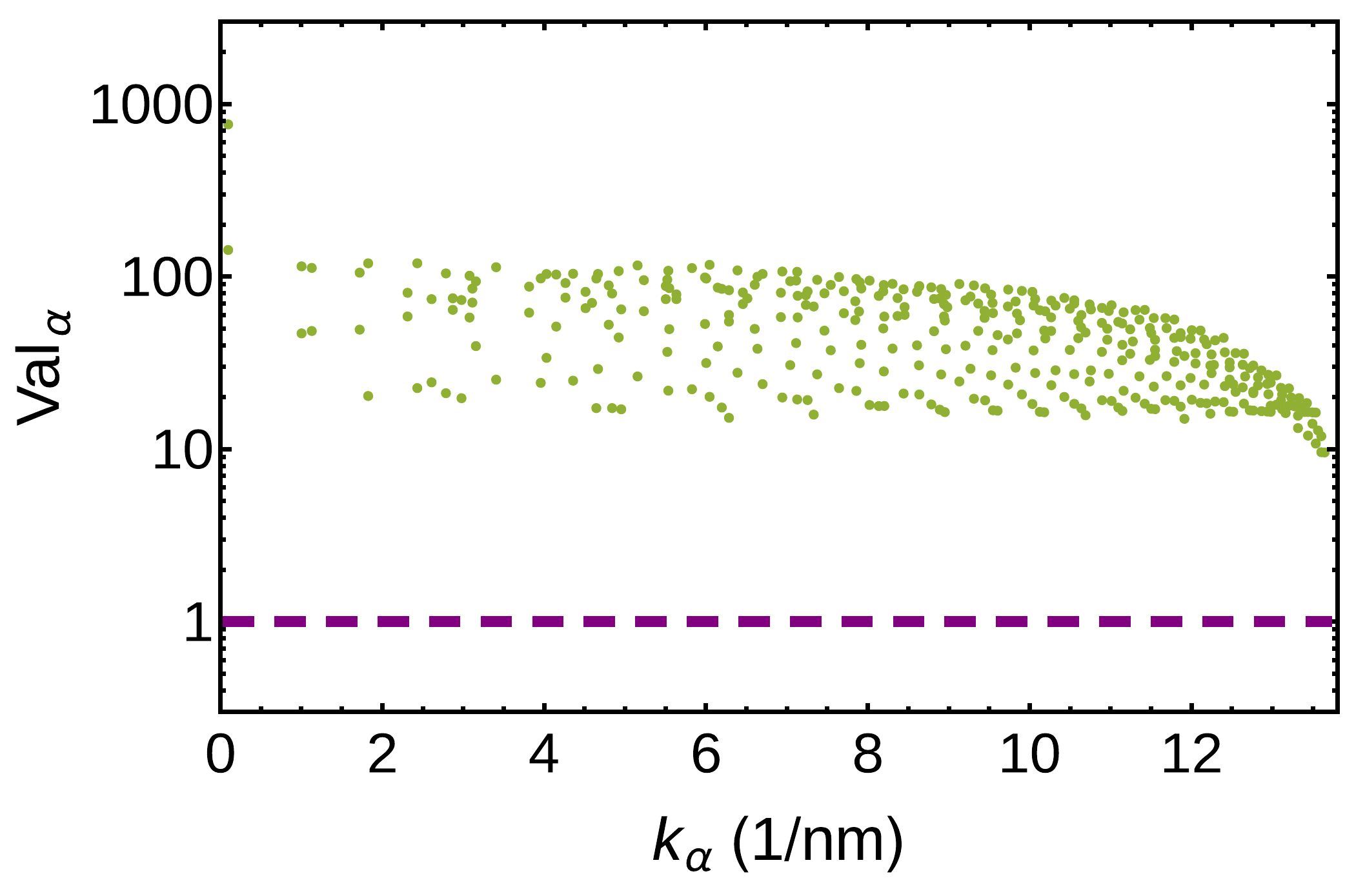}}
   \end{center}
   \caption{(a) The validity of Fermi's golden golden rule, used to
	    estimate surface roughness (SR) relaxation times (see
	    Eq.~\ref{validitySR}) is checked for a Cu nanowire with
	    length $L_z = 100$~nm, lattice constant
	    $a_\textnormal{Cu} \approx 0.361$~nm and electron density
	    $n_e \approx 8.469 \times 10^{22}$~cm${}^{-3}$. The wire
	    width and height are taken to be $L_x = L_y = 10$~nm and
	    the SR parameters are chosen to be $\Delta = a_\textnormal{Cu}$
	    and $\Lambda = 3a_\textnormal{Cu}$. The barrier height is
	    estimated from $V = E_\textnormal{\tiny F} + W$, evaluated with
	    bulk values for work function and Fermi energy: $W = 4.5$~eV,
	    $E_\textnormal{\tiny F} = 7$~eV. (b) Same as (a), but with
	    $L_x=L_y \approx 5$~nm, $\Delta = a_\textnormal{Cu}/2$ and
	    $\Lambda = 6 a_\textnormal{Cu}$
            }
   \label{FigSRValidity}
\end{figure}

We represent surface roughness by a scattering potential being the
difference between the smooth potential well of an ideal wire and
potential well profile that is shifted due to the surface roughness.
The deviation of the rough surface with respect to the ideal smooth
surface is characterized by a surface function $\Deltabd{}(x/y,z)$,
the subscript ``bd'' referring to the nanowire boundary surface.
We proceed with an analysis for the $x=0$ surface below, for which the
potential is given by:
\begin{equation}
   V^\textnormal{\tiny SR}_{x = 0}(\vecr) \equiv
   H_0^\textrm{pot.} \lef x - \Delta_{x = 0}(y, z), y , z \rig -
   H_0^\textrm{pot.} \lef x , y , z \rig \approx
   U \left[
           \theta \lef x - \Delta_{x = 0}(y, z) \rig - \theta \lef x \rig
     \right],
\end{equation}
where $H_0^\textrm{pot.}$ is the potential energy part of the
Hamiltonian and a finite potential well with barrier height $U$ along
the confinement directions is considered, so as to ensure a well-defined
shift of the potentials.
Because the whole Hamiltonian gets shifted over a distance
$\Delta_{x = 0}(y, z)$, there also appears a non-zero contribution
from near the boundary opposite of the $x = 0$ surface. This, however,
should not be considered part of the matrix element as it is just
an artifact of the notation. Following the same procedure as before,
we calculate the matrix elements,
\begin{equation}
   \mathcal{M}_{if}^{\textnormal{\tiny SR}(x = 0)} \approx
   \frac{U}{L_z} \psi^*_i(x = 0) \psi_f(x = 0)
   \mkern-5mu \int \limits_{-\infty}^{+\infty} \mkern-7mu \deriv y \;
   \psi^*_i(y) \psi_f(y)
   \mkern-11mu \int \limits_{-L_z/2}^{+L_z/2} \mkern-13mu \deriv z \;
   \Delta_{x=0}(y, z) \e^{-\imu (k_i - k_f) z},
\end{equation}
The matrix element is expanded linearly in the surface function around
$\Delta_{x=0} =0$ to simplify the averaging procedure. The functional
integration which, in principle, is still part of the averaging
procedure, reduces now to an integration over surface functions
(that are considered to be independent) rather than over
scattering potential $V$:
\begin{equation}
   \int \mkern-5mu \delta V(\vecr) \; g \left[ V(\vecr) \right] =
   \mkern-3mu \int \mkern-5mu \delta \Delta_{x = 0} \;
   g \left[ \Delta_{x = 0} \right]
   \cdots
   \mkern-3mu \int \mkern-5mu \delta \Delta_{y = L_y} \;
   g \left[ \Delta_{y=L_y} \right].
\end{equation}
In turn, the functional integration over a product of surface
roughness functions is assumed to result in a multivariate normal
distribution, not specifying the underlying distribution function $g$.
The following identities will be used:
\begin{align}
   & \left\langle \Deltabd{} \right\rangle_V = 0, \qquad
     \left\langle \Deltabd{1} \Deltabd{2} \right\rangle_V =
     \delta_{\textnormal{bd}_1, \textnormal{bd}_2}
     \Delta^2 \rho(\vecr_1, \vecr_2), \qquad
     \left\langle \Deltabd{1} \Deltabd{2} \Deltabd{3} \right\rangle_V = 0, \\ \notag
   & \left\langle
                 \Deltabd{1} \Deltabd{2} \Deltabd{3} \Deltabd{4}
     \right\rangle_V = \Delta^4
     \left[
           \delta_{\textnormal{bd}_1, \textnormal{bd}_2}
           \delta_{\textnormal{bd}_3, \textnormal{bd}_4}
           \rho(\vecr_1, \vecr_2) \rho(\vecr_3,\vecr_4) +
           (2 \leftrightarrow 3) + (2 \leftrightarrow 4)
     \right].
\end{align}
The last line relies on the well-known Wick  theorem \cite{wick1950evaluation}
(or Isserlis theorem \cite{isserlis1918formula}) and establishes various
contributions from all possible permutations of the boundary indices.
We consider surface roughness functions with mean equal to zero, standard
deviation $\Delta$ and a Gaussian autocorrelation function with
correlation length $\Lambda$:
\begin{equation}
   \rho \lef \vecr_1, \vecr_2 \rig \equiv
   \e^{-(\vecr_1 - \vecr_2)^2 / (\Lambda^2/2)},
\end{equation}
which leads to the following results for the second- and third-order contributions:
\begin{align}
   \mathcal{O} \lef V^2 \rig
   & = \sum_\textnormal{bd}
       \left\langle
       \left|
             \mathcal{M}_{if}^{\textnormal{\tiny SR}(\textnormal{bd})}
       \right|^2
       \right\rangle_V, \qquad \mathcal{O} \lef V^3 \rig \approx 0, \\ \notag
   \left\langle
   \left| \mathcal{M}_{if}^{\textnormal{\tiny SR}(x=0)} \right|^2
   \right\rangle_V
   & \approx \mkern-3mu \int \mkern-5mu \deriv z_1 
     \mkern-3mu \int \mkern-5mu \deriv z_2 \;
     \frac{\e^{\imu (k_i-k_f) (z_1 - z_2)}}{L_z^2}
     \rho \lef z_1, z_2 \rig \lef V \Delta \rig^2
     \left| \psi_i(x=0) \right|^2 \left| \psi_f(x=0) \right|^2 \;
     \tensor*[_{y_1}^{y_2}]{C}{_{i f}^{f i}},
\end{align}
where $\tensor*[_{y_1}^{y_2}]{C}{_{i f}^{f i}}$ is defined by
\begin{align}
   \tensor*[_{y_1}^{y_2}]{C}{_{ab}^{cd}}
   & \equiv \mkern-5mu \int\limits_{-\infty}^{+\infty}
     \mkern-7mu \deriv y_1 \; \psi^*_a \lef y_1 \rig \psi_b \lef y_1 \rig
     \mkern-5mu \int\limits_{-\infty}^{+\infty} \mkern-7mu \deriv y_2 \;
     \psi^*_c \lef y_2 \rig \psi_d \lef y_2 \rig
     \e^{- (y_1 - y_2)^2/(\Lambda^2/2)},
\end{align}
and analogous short-cuts apply to the integrals over $x$.
We develop the two terms containing the fourth-order contributions:
\begin{align}
   & \left\langle
     \left|
           \sum\limits_{\alpha} 
           \lef \frac{-\imu L_z}{2 B_\alpha k_\alpha^+} \rig
           \overline{\mathcal{M}_{f\alpha i}^\textnormal{\tiny SR}}
     \right|^2 + 2 \mathcal{R}
     \left[
           \sum\limits_{\alpha,\alpha'}
           \lef \frac{-\imu L_z}{2 B_\alpha k_\alpha^+} \rig
           \lef \frac{-\imu L_z}{2 B_{\alpha'} k_{\alpha'}^+} \rig
           \overline{\mathcal{M}_{f\alpha\alpha' i}^\textnormal{\tiny SR}}
           \mathcal{M}_{if}^\textnormal{\tiny SR}
     \right]
     \right\rangle_V \\ \notag
   & \; = \sum_{\alpha, \alpha'} \frac{L_z}{2 B_\alpha k_\alpha^+}
     \frac{L_z}{2 B_{\alpha'} k_{\alpha'}^+}
     \sum_{\textnormal{bd}_{i(1,\ldots,4)}} \mkern-15mu
     \lef
         \delta_{\textnormal{bd}_1, \textnormal{bd}_2}
         \delta_{\textnormal{bd}_3, \textnormal{bd}_4} +
         \textnormal{permutations}
     \rig \\ \notag
   & \; \quad \qquad \qquad \times
     \left[
           \left\langle
           \overline{\mathcal{M}_{f\alpha i}^{\textnormal{\tiny SR}
                     (\textnormal{bd}_1, \textnormal{bd}_2)}} \,
           \overline{\mathcal{M}_{i\alpha' f}^{\textnormal{\tiny SR}
                     (\textnormal{bd}_3, \textnormal{bd}_4)}}
           \right\rangle_V - 2 \mathcal{R}
           \lef \left\langle
           \overline{\mathcal{M}_{f\alpha\alpha' i}^{\textnormal{\tiny SR}
                     (\textnormal{bd}_1, \textnormal{bd}_2,
                      \textnormal{bd}_3)}}
           \mathcal{M}_{if}^{\textnormal{\tiny SR}(\textnormal{bd}_4)}
           \right\rangle_V \rig
     \right],
\end{align}
with for example:
\begin{align}
   & \left\langle
     \overline{\mathcal{M}_{f\alpha i}^{\textnormal{\tiny SR}(x=0,x=0)}} \,
     \overline{\mathcal{M}_{i\alpha' f}^{\textnormal{\tiny SR}(x=0,x=0)}}
     \right\rangle_V \\ \notag
   & \; \approx \mkern-3mu \int \mkern-5mu \deriv z_1 \mkern-3mu \int
     \mkern-5mu \deriv z_2 \mkern-3mu \int \mkern-5mu \deriv z_3
     \mkern-3mu \int \mkern-5mu \deriv z_4 \;
     \frac{\e^{\imu k_f (z_4 - z_1) - \imu k_\alpha^+ |z_1 - z_2| +
           \imu k_i (z_2 - z_3) + \imu k_{\alpha'}^+ |z_3 - z_4|}}
          {L_z^4}
     \left[
           \rho \lef z_1, z_2 \rig \rho \lef z_3, z_4 \rig +
           \textnormal{perm.}
     \right] \\ \notag
   & \; \quad \times \lef V \Delta \rig^4
     \left| \psi_f \lef x = 0 \rig \right|^2
     \left| \psi_\alpha \lef x = 0 \rig \right|^2
     \left| \psi_i \lef x = 0 \rig \right|^2
     \left| \psi_{\alpha'} \lef x = 0 \rig \right|^2 \;
     \tensor*[_{y_1}^{y_2}]{C}{_{f \alpha}^{\alpha i}} \;
     \tensor*[_{y_3}^{y_4}]{C}{_{i \alpha'}^{\alpha' f}}.
\end{align}
We treat the remaining integrals along the transverse
and transport directions by approximating the wave functions
by the infinite potential well solutions (see appendix
\ref{appendixGaussian}) to obtain the following validity
criterion for roughness at the $x=0$ boundary surface,
ignoring corrections of order one:
\begin{align} \label{validitySR}
   \prod_{\alpha = i, f} \textnormal{Val}_\alpha &\ll 1, \\
   \textnormal{Val}_\alpha &\equiv \sum_\beta \sqrt{\frac{\pi}{2}}
   \frac{\Lambda}{\sqrt{L_{y} L_z}}
   \frac{V \Delta \left| \psi_\beta (x = 0) \right|^2}
        {\left| 2 B_\alpha k_\alpha / L_z \right|}
   \theta^x_{\alpha \beta} ( \sqrt{8}/\Lambda ).
\end{align}
$\theta^x_{\alpha \beta} ( \sqrt{8}/\Lambda )$ represents
a constraint on the difference of wave vectors of Fermi
level states $\mid \! \alpha \rangle$ and $\mid \! \beta \rangle$,
one of them being the initial or final state and the other
an intermediate state. The difference should be less than
$\Delta k$, including for the standing wave vector along
the transverse direction parallel to the $x=0$ boundary plane:
\begin{align} \label{eqSelectionGaussian}
    \theta^x_{\alpha \beta}( \Delta k) &\equiv
	\begin{cases}
	  1	& \textnormal{if} \quad \left| \pi n_{\beta \, y} /
		L_y - \pi n_{\alpha \, y} / L_y \right| < \Delta k
		\textnormal{ and } \left| k_\beta - k_\alpha \right| < \Delta k \\
	  0	& \textnormal{else}
	\end{cases}.
\end{align}
Intermediate states only contribute substantially to the
validity criterion when the constraint on the wave vector,
as defined in Eq.~\ref{eqSelectionGaussian}, is met for
$\Delta k = \sqrt{8}/\Lambda$ due to the Gaussian surface
roughness profile which otherwise exponentially suppresses
their contribution, as shown in appendix~\ref{appendixGaussian}.
Note that the criterion is again length dependent, although
the lowest-order scattering rate results in length independent
transport properties. The above criterion is evaluated for
two nanowires with different surface roughness properties,
the outcome of which is depicted in Fig.~\ref{FigSRValidity}.

\section{Discussion}
\label{sectionDiscussion}
Thorough investigation of the three scattering mechanisms discussed
in section~\ref{sectionMetalNW} reveals that Fermi's golden
rule may fail to generate reliable scattering rates for realistic
scattering potentials, as can be concluded in particular for electrons
in metallic nanowires with a single parabolic conduction band from
Figs.~\ref{FigImpValidity}, \ref{FigGBValidity} and \ref{FigSRValidity}.
The validity can be verified easily in this case by evaluating the
validity criteria and depends on the strength of the scattering
potential, both in energy and size, and the allowed couplings
to intermediate states. The corresponding results clearly indicate
that higher-order effects, superseding the second-order perturbation
treatment provided by Fermi's golden rule may come into play and
become dominant.

For impurity scattering, the interaction strength, represented by
the coefficient that precedes the delta function in the simple model,
may be relatively small in practice but the higher-order contributions
allow for all Fermi level states as intermediate states, as can be
seen in Eq.~\ref{ImpValidityEq}. Hence, the validity of the golden rule
scattering rate may be violated for relatively small interaction strengths.

For grain boundaries, the barrier strength associated with a single
boundary plane is typically larger than that of a local impurity,
but the allowed intermediate states are much more restricted. A
perpendicular orientation of grain boundary planes restricts
the intermediate states to states with equal subband indices,
thus avoiding large summations over $\alpha$ in Eq.~\ref{critGB}.
However, violation of the scattering rate validity can be reached
when the grain boundary strength or density is large enough,
which applies most pronouncedly to electron states with low
transport momentum. This can be understood physically as these
states get trapped most easily in between grain boundaries through
higher-order interactions and effectively become bound states,
not contributing to the drive current.

Finally, the golden rule scattering rates obtained for surface
roughness scattering appear to be most error prone of all
three mechanisms. The scattering potential associated with
surface roughness is proportional to the energy barrier height
outside the metallic wire, which substantially exceeds any
internal barriers. Even though the coupling to intermediate
states is limited due to the scattering wave vector difference
being limited by the inverse of the surface roughness correlation
length, the higher-order processes violate the validity criteria
for realistic parameters easily. Truly, the validity criteria in
Eq.~\ref{validitySR} are obtained without rigorously computing
the different diagrams and for a linear expansion of the matrix
elements for small roughness sizes (which can be improved upon
\cite{lizzit2014new,moors2015modeling}). However, since the fourth
order contribution exceeds the lowest-order scattering rate
by many orders of magnitude, the discussion about the application of
perturbation theory to surface roughness scattering remains open,
not only in relation with metallic nanowires but also in other
areas, e.g. when rough edge scattering in graphene ribbons
\cite{fischetti2011empirical} is explored, or when comparison with
non-perturbative approaches, based on non-equilibrium Green
functions, comes into play \cite{niquet2014quantum}.

\section{Conclusion}
\label{sectionConclusion}
We have developed a framework to derive criteria that can be used to
check the validity of Fermi's golden rule scattering rates
systematically, such that their applicability for transport modeling
in condensed matter systems can be easily verified. This framework
includes an ensemble average over scattering potential profiles, which
can be formally represented as a functional integral, leading to general
validity criteria that depend on the crucial system parameters, e.g.
system size and electron effective mass, and statistical properties
of the scattering sources, e.g. impurity strength or surface
roughness standard deviation. One can derive a criterion for each
higher-order term in the perturbation expansion of the scattering rates.

With the presented framework, we were able to derive simple and
general validity criteria for localized impurity scattering
(Eq.~\ref{ImpValidityEq}), grain boundary scattering (Eq.~\ref{critGB})
and surface roughness scattering (Eq.~\ref{validitySR}) in metallic
nanowires with a single parabolic conduction band, based on the third-
and fourth-order corrections to Fermi's golden rule. The fourth-order
correction leads to the strongest validity constraint with which we are
able to identify the different aspects having an impact on the validity
such as the scattering source strength and its particular coupling
properties to intermediate states. All these aspects are uniquely
determined for each type of scattering potential and play a crucial
role in a rigorous validity analysis of the scattering rates and cannot
be taken into account through a merely qualitative analysis of the
higher-order corrections, hence confirming the importance and
advantages of this type of general criteria. A derivation of
validity criteria for nanowires with more general band structures
and other (inelastic) scattering mechanisms remains open for
future work.

\bibliography{Validity_Moors_Final}{}

\appendix

\section{Feynman diagrams}
\label{appendixFeynman}
\begin{figure*}[tb]
 \begin{center}
      \subfigure{
      \includegraphics[height=0.15\linewidth]{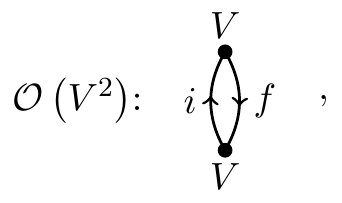}}
      \subfigure{
      \includegraphics[height=0.15\linewidth]{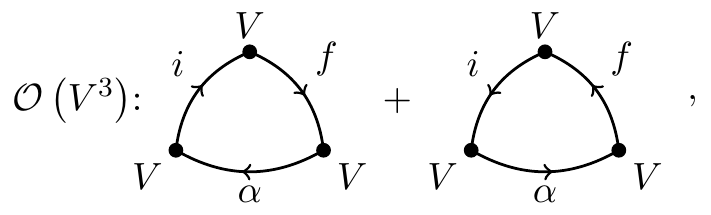}}
      \subfigure{
      \includegraphics[height=0.15\linewidth]{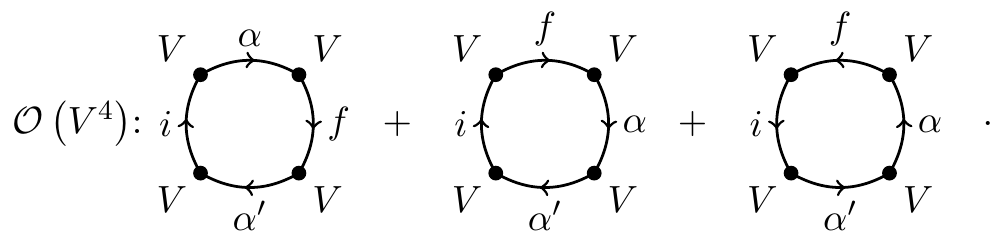}}
 \end{center}
\caption{
         Feynman diagrams corresponding to the
         $\mathcal{O}(V^2)$, $\mathcal{O}(V^3)$ and $\mathcal{O}(V^4)$
         contributions to the scattering rate.
         }
\label{FeynmanDiagram}
\end{figure*}

\begin{figure*}[tb]
\begin{center}
      \subfigure{
      \includegraphics[height=0.17\linewidth]{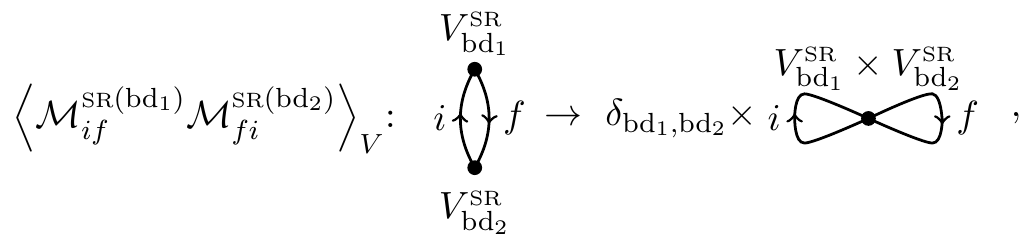}}
      \subfigure{
      \includegraphics[height=0.17\linewidth]{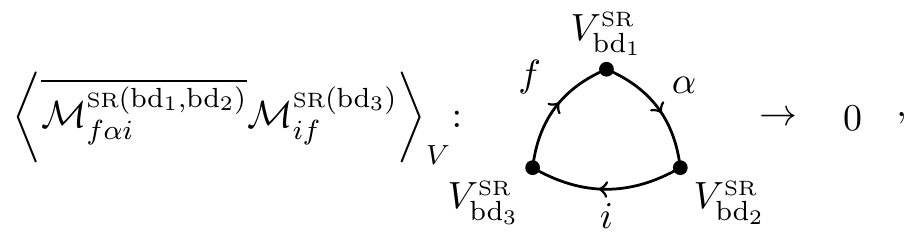}}
\end{center}
\caption{
         Diagrammatic representation of the terms that
         contribute to the validity criteria for surface
         roughness, as developed in section~\ref{sectionSR}
         (part I).
         }
\label{SRFeynmanDiagram}
\end{figure*}

\begin{figure*}[tb]
\begin{center}
      \subfigure{
      \includegraphics[width=0.8\linewidth]{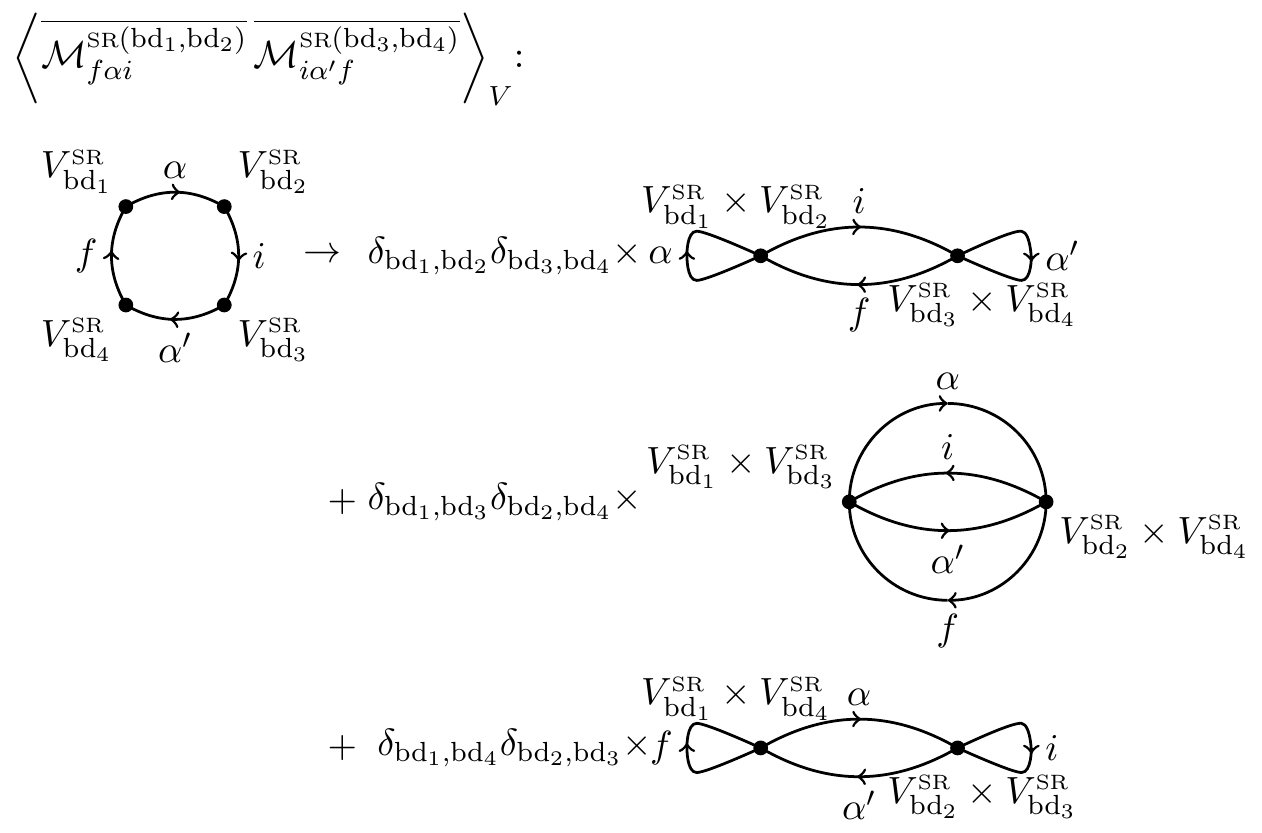}}
      \subfigure{
      \includegraphics[width=0.8\linewidth]{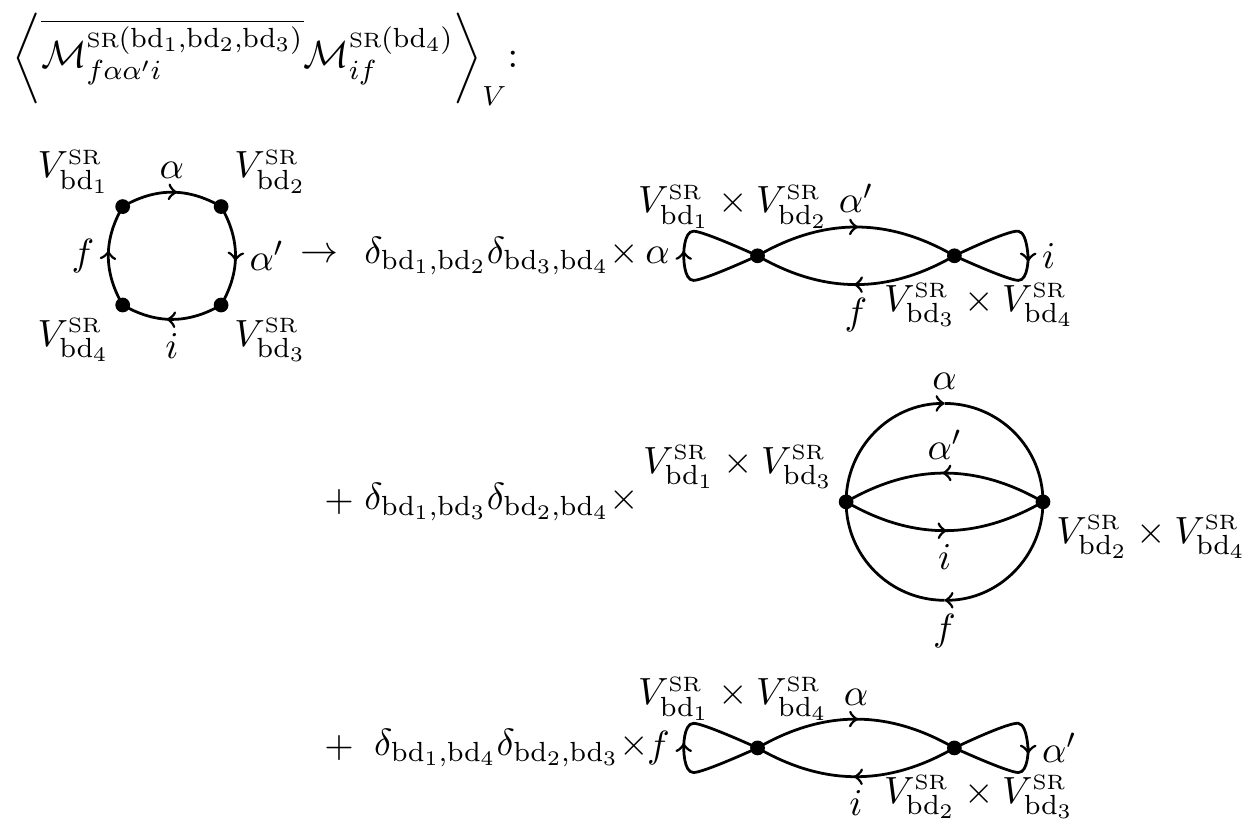}}
\end{center}
\caption{
         Diagrammatic
         representation of the terms that contribute to the validity
         criteria for surface roughness, as developed in
         section~\ref{sectionSR} (part II).
         }
\label{SRFeynmanDiagram2}
\end{figure*}

Scattering contributions can be evaluated up to arbitrary order with
the help of Feynman diagrams. Below, a set of simple diagrammatic rules
is summarized together with a proper diagrammatic notation.
\begin{itemize}
   \item Diagrammatic rules
   \begin{itemize}
      \item Draw a number of vertices equal to the order of $V$ under
            consideration.
      \item Draw all combinations of directed arrows between two different
            vertices (no arrow coming back to the same vertex), such that each
            vertex has a single incoming arrow and a single outgoing arrow,
            all vertices being connected through a single loop.
            Two arrows should be labeled $i$ and $f$ so as to represent
            respectively the initial and final state.
      \item If an identical diagram arises from reversing all arrows and
            renaming all labels but $i$ and $f$, it should be discarded.
   \end{itemize}
   \item Contributions to the scattering rates
   \begin{itemize}
      \item All labels unequal to $i$ or $f$ represent electron states at
            the Fermi level (possibly equal to state
            $\mid \! i \rangle$ or $\mid \! f \rangle$) and add a factor
            $L_z  / 2 B_\textnormal{label} k_\textnormal{label}$ to the
            scattering rate.
      \item Add a factor
            $(-\imu)^\textnormal{``number of $V$ between $f$ and $i$'' - 1}
            \times (\imu)^\textnormal{``number of $V$ between $i$ and 
            $f$'' - 1}$.
      \item For each vertex, add a factor
            $\langle \textnormal{label incoming arrow} \mid V \mid
            \textnormal{label outgoing arrow} \rangle$
            to the scattering rate.
      \item Sum over all Fermi level states for each label not representing
            the initial or final state and correct the matrix elements by
            changing the differences in $z$-positions $\Delta z$ that
            multiply the $k_\textnormal{label}$ wave vectors by their
            absolute value and fixing the sign such that
            $+\imu|k_\textnormal{label} | | \Delta z |$ appears in the
            exponential wave function along the $z$-direction
            if the label appears in between $i$ and $f$ and
            $-\imu |k_\textnormal{label}| | \Delta z |$ if the label appears in
            between $f$ and $i$ on the oriented loop running between the vertices.
      \item Finally, multiply by $2\pi / \hbar \times \delta(E_i - E_f)$.
   \end{itemize}
\end{itemize}
All diagrams of the second up to the fourth-order contribution
to the scattering rate are shown in Fig.~\ref{FeynmanDiagram}.
Higher order diagrams can be calculated by applying Wick's theorem
(see Fig.~\ref{SRFeynmanDiagram} and Fig.~\ref{SRFeynmanDiagram2}),
which was used to obtain the validity criteria for surface roughness
scattering in section~\ref{sectionSR}.

\setlength{\unitlength}{1.5cm}

\section{Gaussian integrals}
\label{appendixGaussian}
The expression $\tensor*[_{y_1}^{y_2}]{C}{_{ab}^{cd}}$ can be
evaluated by approximating the wave functions by the infinite
potential well solutions (analogously for
$\tensor*[_{x_1}^{x_2}]{C}{_{ab}^{cd}}$), yielding
\begin{align}
   \tensor*[_{y_1}^{y_2}]{C}{_{ab}^{cd}}
   & \approx \sqrt{\frac{\pi}{2}} \frac{\Lambda}{L_y} \frac{1}{2}
     \left\{
            \e^{-(n_c^y - n_d^y)^2 \pi^2 \Lambda^2 / 8 L_y^2}
            \left[
                  \delta_{|n_a^y - n_b^y|, |n_c^y - n_d^y|}
                  \lef 1 + \delta_{n_a^y, n_b^y} \rig -
                  \delta_{n_a^y + n_b^y, |n_c^y - n_d^y|}
            \right]
     \right. \\ \notag
   & \quad \qquad \qquad \quad
     \left.
           + \e^{-(n_c^y + n_d^y)^2 \pi^2 \Lambda^2 / 8 L_y^2}
           \lef
               \delta_{n_a^y + n_b^y, n_c^y + n_d^y} -
               \delta_{|n_a^y - n_b^y|, n_c^y + n_d^y}i
           \rig
     \right\}.
\end{align}
Calculation of the remaining integrals along the transport
direction yields the second-order contribution as follows:
\begin{align}
   \mkern-13mu \int\limits_{-L_z/2}^{+L_z/2} \mkern-15mu \deriv z_1
   \mkern-10mu \int\limits_{-L_z/2}^{+L_z/2} \mkern-15mu \deriv z_2 \;
   \frac{\e^{\imu (k_i-k_f) (z_1 - z_2)}}
        {L_z^2} \rho \lef z_1, z_2 \rig \approx \sqrt{\frac{\pi}{2}}
   \frac{\Lambda}{L_z} \e^{-(k_i - k_f)^2 \Lambda^2 / 8}.
\end{align}
Evaluation of the fourth-order contribution requires a careful
treatment of the position differences appearing in the wave
functions. We get for example:
\begin{align}
   & \int \limits_{-L_z/2}^{+L_z/2} \mkern-15mu \deriv z_1 \mkern-10mu
     \int \limits_{-L_z/2}^{+L_z/2} \mkern-15mu \deriv z_2 \mkern-10mu
     \int \limits_{-L_z/2}^{+L_z/2} \mkern-15mu \deriv z_3 \mkern-10mu
     \int \limits_{-L_z/2}^{+L_z/2} \mkern-15mu \deriv z_4 \;
     \frac{\e^{\imu k_f (z_4 - z_1) - \imu k_\alpha^+ |z_1 - z_2| +
           \imu k_i (z_2 - z_3) + \imu k_{\alpha'}^+ |z_3 - z_4|}}{L_z^4}
     \rho \lef z_1, z_2 \rig \rho \lef z_3, z_4 \rig \\ \notag
   & \; \approx \frac{\pi}{2} \lef \frac{\Lambda}{L_z} \rig^2
     \frac{2 - 2\cos\left[ \lef k_f-k_i \rig L_z \right]}
          {\lef k_f - k_i \rig^2 L_z^2}
     \sum_{\pm, \pm'} \frac{\e^{-(k_i \pm k_\alpha^+)^2 \Lambda^2 / 8}}{2}
     \frac{\e^{-(k_f \pm' k_{\alpha'}^+)^2 \Lambda^2 / 8}}{2}, \\
   & \int \limits_{-L_z/2}^{+L_z/2} \mkern-15mu \deriv z_1 \mkern-10mu
     \int \limits_{-L_z/2}^{+L_z/2} \mkern-15mu \deriv z_2 \mkern-10mu
     \int \limits_{-L_z/2}^{+L_z/2} \mkern-15mu \deriv z_3 \mkern-10mu
     \int \limits_{-L_z/2}^{+L_z/2} \mkern-15mu \deriv z_4 \;
     \frac{\e^{\imu k_f (z_4 - z_1) - \imu k_\alpha^+ |z_1 - z_2| - 
           \imu k_{\alpha'}^+ |z_2 - z_3| + \imu k_i (z_3 - z_4)}}{L_z^4}
     \rho \lef z_1, z_2 \rig \rho \lef z_3, z_4 \rig \\ \notag
   & \; \approx \frac{\pi}{2} \lef \frac{\Lambda}{L_z} \rig^2
     \e^{-(k_f - k_i)^2 \Lambda^2/8}
     \sum_{\pm, \pm'} \frac{\e^{-(k_f \pm k_\alpha^+)^2 \Lambda^2/8}}{2}
     \frac{\delta_{k_{\alpha'}^+, \pm' k_f}}{2}.
\end{align}

\end{document}